\documentclass[11pt,twoside]{article}

\usepackage{asp2004}
\usepackage{epsf}
\usepackage{psfig}
\usepackage{lscape}

\begin{document}
\title{Where do $z\approx$~2 Submillimeter-Emitting Galaxies Lie On the Black-Hole--Spheroid Mass Plane?}
\author{D.~M.~Alexander}   %%% Fill in author names
\affil{Department of Physics, Durham University, Durham, DH1 3LE, UK}    %%% Fill in author affiliations

\begin{abstract} %%% Abstract to run on from here.
Submillimeter-emitting galaxies (SMGs) are $z\approx$~2 bolometrically
luminous systems hosting energetic starburst and AGN activity. SMGs
may represent a rapid growth phase that every massive galaxy undergoes
before lying on the well-established black-hole--spheroid mass
relationship in the local Universe. Here we briefly discuss our recent
results from Alexander et~al. (2008) where we estimated the masses of
the black holes in SMGs using the black-hole virial mass estimator,
finding $M_{\rm BH}\approx6\times10^{7}$~$M_{\odot}$ for typical
SMGs. We show that the black-hole--spheroid mass ratio for SMGs at
$z\approx$~2 was suggestively below that found for massive galaxies in
the local Universe and more than an order of magnitude below the
black-hole--spheroid mass ratio estimated for $z\approx$~2 quasars and
radio galaxies. We demonstrate that SMGs and their progeny cannot lie
on the elevated $z\approx$~2 black-hole--spheroid mass relationship of
quasars--radio galaxies without overproducing the space density of the
most massive black holes ($M_{\rm BH}\approx10^9$~$M_{\odot}$), unless
the galaxy spheroid of SMGs is an order of magnitude lower than that
typically assumed ($M_{\rm SPH}\approx10^{10}$~$M_{\odot}$). We also
show that the relative black-hole--spheroid growth rates of typical
SMGs appear to be insufficient to significantly increase the
black-hole--spheroid mass ratio, without requiring long duty cycles
($>10^9$~years), and argue that a more AGN-dominated phase (e.g.,\ an
optically bright quasar) is required to significantly move SMGs (and
their progeny) up the black-hole--spheroid mass plane.

\end{abstract}

%%% MAIN BODY OF TEXT GOES HERE. CONSULT "INSTRUCTIONS FOR AUTHORS USING
%%% LATEX2E MARKUP", SECTIONS 2.3-2.6 FOR HELP WITH EQUATIONS, FIGURES,
%%% AND TABLES.

\section{Introduction}   %%% Top level section head (remove "%" symbol)

Due to the negative $K$-correction for infrared-luminous galaxies at
$z>1$, submillimeter/millimeter surveys select the most bolometrically
luminous systems in the Universe (e.g.,\ Blain et~al. 2002). After
intense multi-wavelength follow-up observations it is now clear that
submillimeter-emitting galaxies (SMGs; $f_{\rm 850\mu m}>4$~mJy) are
gas-rich massive galaxies at $z\approx$~2--4 hosting energetic
starburst and AGN activity (e.g.,\ \citealt{smail02};
\citealt{alex05a}; \citealt{chap05}); the stellar--dynamical and CO
gas masses of these galaxies are typically
$\approx10^{11}$~$M_{\odot}$ and $\approx3\times10^{10}$~$M_{\odot}$,
respectively (e.g.,\ \citealt{swin04}; \citealt{borys05};
\citealt{greve05}). The compactness and high inferred gas density of
the CO emission from SMGs suggests that the CO dynamics trace the mass
of the galaxy spheroid (e.g.,\ \citealt{bouche07}). It is commonly
believed that SMGs represent a major-merger induced growth phase that
causes rapid black-hole--spheroid growth, and it is possible that
every massive galaxy ($>$~1--3~$L_{\rm *}$) in the local Universe
underwent at least one submillimeter-bright phase at some time in the
distant past (e.g.,\ \citealt{swin06}).

The close relationship between supermassive black-hole (SMBH) and
spheroid mass--luminosity--velocity dispersion in the local Universe
(e.g.,\ \citealt{gebh00}; \citealt{haring04}) suggests that the growth
of SMBHs and galaxy spheroids was concordant. The potentially rapid
growth rates of the SMBH and host galaxy in SMGs could cause these
systems to deviate significantly from the locally defined
black-hole--spheroid mass relationship, if the growth of one component
is much faster than the other component. It is therefore of interest
to constrain the SMBH masses and relative black-hole--host galaxy
growth in SMGs to give insight into the formation and evolution of
today's massive galaxies.

Using X-ray luminosity derived black-hole masses and photometrically
determined stellar masses, \cite{borys05} showed that SMGs lie more
than an order of magnitude below the locally defined
black-hole--spheroid mass relationship, under the assumption of
Eddington-limited SMBH accretion. While \cite{borys05} provided useful
first-order constraints, it is neccessary to determine the SMBH masses
of SMGs without having to assume the Eddington ratio of the SMBH.
Here we present recent results from \cite{alex08}, where we used the
virial SMBH mass estimator to calculate SMBH masses for a handful of
SMGs with detected broad emission lines. These results are used to
constrain the relative black-hole--spheroid mass ratio of SMGs and to
explore how quickly the SMBH and host galaxy is growing in these
systems.

\section{Weighing the Black Holes in SMGs}

Black-hole masses ($M_{\rm BH}$) have been directly measured for a
small number of galaxies in the local Universe, on the basis of the
velocity dispersion of stars/gas in the vicinity of the SMBH (e.g.,\
\citealt{gebh00}). Black holes cannot be ``weighed'' in the same way
at high redshift due to poorer spatial resolution and lower
signal-to-noise ratio data.  However, the well-established virial SMBH
mass estimator, which works on the assumption that the broad-line
regions (BLRs) in AGNs are virialized, provides an apparently
reliable, if indirect, measurement of SMBH masses in high-$z$ AGNs
(e.g.,\ \citealt{kaspi00}).

The virial SMBH mass estimator is somewhat restrictive for SMGs since
the majority of the sources are heavily obscured (e.g.,\
\citealt{alex05b}; \citealt{chap05}). However, the identification of
broad H$\alpha$/H$\beta$ emission from a small number of SMGs
(\citealt{swin04}; \citealt{takata06}) provides the opportunity to
estimate SMBH masses for a handful of sources so far. In \cite{alex08}
we used the virial estimator of \cite{greene05}, which calculates
$M_{\rm BH}$ solely on the properties of the H$\alpha$/H$\beta$
emission line and reduces potential uncertainties on the luminosity of
the AGN (e.g.,\ contaminating emission from the host galaxy or an
accretion-related jet) when compared to other estimators. Taking the
six SMGs from \cite{swin04} and \cite{takata06} with broad emission
lines (FWHM$>1000$~km~s$^{-1}$) that appear to be intrinsic to the
central source (i.e., unrelated to AGN outflows), we find relatively
modest SMBH masses [average $M_{\rm
BH}\approx(1-3)\times10^{8}$~$M_{\odot}$, depending upon the assumed
geometry of the broad-line region; see \S3.2 in
\citealt{alex08}]. Broad-line SMGs typically have relatively narrow
emission lines when compared to the optically bright quasar population
($\approx$~2500~km~s$^{-1}$ vs typically $\approx$~5000~km~s$^{-1}$),
which is a major factor that leads to the comparatively low SMBH
masses. Under the assumption that X-ray obscured SMGs have essentially
the same AGN properties as the broad-line SMGs but are obscured due to
the presence of an optically and geometrically thick torus aligned
along the line of sight (i.e.,\ the Unified AGN model), we can
calculate the Eddington ratio (defined as $L_{\rm AGN}/L_{\rm Edd}$;
i.e.,\ $\eta$) of the broad-line SMGs and derive SMBH masses for the
X-ray obscured SMGs utilising the Eddington-limited SMBH mass
constraints in \cite{alex05a}. We find a typical Eddington ratio for
the broad-line SMGs of $\eta\approx0.2$, leading to $M_{\rm
BH}\approx6\times10^{7}$~$M_{\odot}$ for a typical X-ray obscured SMG;
see \S3.4 in \cite{alex08} for further justification.

Due to a number of poorly constrained assumptions that need to be made
to calculate Eddington ratios (i.e.,\ the conversion from X-ray
luminosity to mass accretion rate, the derivation of SMBH masses, and
the calculation of the Eddington luminosity limit), the uncertainties
on any estimation of an Eddington ratio are large, particularly at
high redshift. However, the primary reason for the derivation of the
Eddington ratios in \cite{alex08} was to provide a way to estimated
SMBHs for the X-ray obscured SMGs, by scaling the X-ray luminosities
of the broad-line SMGs to those of the X-ray obscured SMGs. In this
approach, we only make the asssumption that the broad-line SMGs are
the unobscured counterparts of the larger X-ray obscured SMG
population, and therefore all of the above uncertainties become
negligible.

Two alternative scenarios for the relationship between broad-line SMGs
and X-ray obscured SMGs are that the broad-line SMGs either represent
a later stage in the evolution of the SMG population or are a higher
SMBH accretion rate phase (e.g.,\ \citealt{alex05b}; \citealt{stev05};
\citealt{copp08}). Under these assumptions, the SMBH masses of the
X-ray obscured SMGs could be higher than those derived here (i.e., up
to the $M_{\rm BH}\approx3\times10^8$~$M_{\odot}$ maximum calculated
for the broad-line SMGs). However, as we show in \S4, it is unlikely
that the SMBHs hosted by {\it typical} SMGs can be this massive. In
any case, given the similarity between the estimated Eddington ratios
of the broad-line SMGs and a handful of obscured ultra-luminous
infrared galaxies (ULIRGs) in the local Universe (the SMBH masses for
the obscured ULIRGs were calculated using the broad Pa$\alpha$
emission line, which is detected in a small fraction of obscured
systems; e.g.,\ \citealt{veil99}; see \S4.1 in \citealt{alex08}), we
favour $\eta\approx$~0.2 and $M_{\rm
BH}\approx6\times10^{7}$~$M_{\odot}$ for a typical X-ray obscured
SMG. See \S4 of \cite{alex08} for evidence that the AGNs found in
$z\approx$~2 SMGs are scaled-up versions of those found in local
ULIRGs.

 \setcounter{figure}{0}
 \begin{figure}[!t]
 \plotone{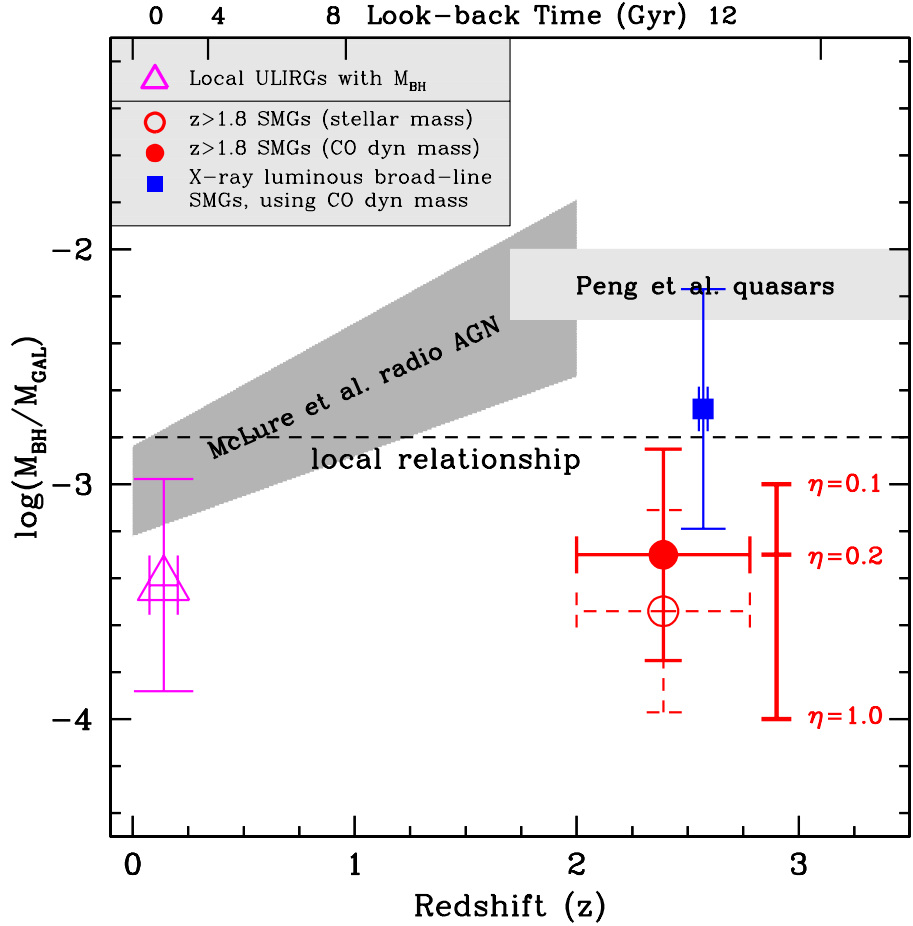}
 \caption{Black-hole--host galaxy (corresponding to the spheroid mass
   for these systems) mass relationship for SMGs and other systems (as
   indicated). SMGs (circles) cannot lie a factor $\approx$~4--6 above
   the local relationship (as found for $z\approx$~2 quasars and radio
   galaxies; \citealt{peng06}; \citealt{mclure06}) without
   overproducing the local black-hole mass density; see \S4 for
   discussion. The solid square indicate where X-ray luminous
   broad-line SMGs would lie, assuming that the average CO dynamical
   mass represents the spheroid mass in these systems, and shows that
   this subset of the broad-line SMG population could be more evolved
   than typical systems. The solid bar indicates how the SMBH masses
   for the SMGs varies depending upon the assumed Eddington ratio
   ($\eta$). This figure is taken from \cite{alex08}; see \S5 of that
   paper for further details.}
 \end{figure}

\section{Exploring the Black-Hole--Spheroid Mass Relationship in SMGs}

By combining the SMBH mass constraints with host-galaxy masses, in
\cite{alex08} we explored the $M_{\rm BH}$--$M_{\rm SPH}$ relationship
for $z\approx2$ SMGs. On the basis of CO spectroscopy and
interferometric imaging, the dynamical masses of SMGs within the
central 2--4~kpc are $M_{\rm DYN}\approx10^{11}$~$M_{\odot}$ (e.g.,\
\citealt{greve05}; \citealt{bouche07}); similar constraints are
derived from $H\alpha$ emission-line widths (\citealt{swin04}). Given
the high inferred gas density of the CO emission in SMGs (comparable
to or higher than the mass density of galaxy spheroids in the local
Universe) it is likely that this dynamical estimate corresponds to the
mass of the spheroid (e.g.,\ \citealt{bouche07}). By comparison, the
average stellar mass derived from ultra-violet--near-IR photometric
data by \cite{borys05} is $2\times10^{11}$~$M_{\odot}$; here we only
consider the six $z>1.8$ SMGs in \cite{borys05} that do not have UV or
near-IR excess emission over that expected from stars, giving a value
$\approx$~2 times lower than estimated by \cite{borys05}. The
stellar-mass estimate is slightly higher than the masses estimated
from CO and H$\alpha$ dynamics, and may be more representative of the
overall mass of the system rather than just the spheroid; however, we
note that uncertainties in photometrically derived stellar masses are
typically a factor of $\approx$~3.

In Fig.~1 we show the average SMBH--host-galaxy mass ratio for the
X-ray obscured SMGs and compare it to that determined for distant
quasars--radio galaxies (e.g.,\ \citealt{mclure06}; \citealt{peng06})
and local obscured ULIRGs with broad Pa$\alpha$ derived SMBH
masses. The SMBHs of the X-ray obscured SMGs are suggestively smaller
than those of comparably massive galaxies in the local Universe, by a
factor of $\approx$~3--7, although given the number of assumptions in
the determination of these quantities this result should not be
considered statistically conclusive. However, we note that we find the
same result for local ULIRGs, providing corroborating evidence that
SMBHs in $z\approx$~2 SMGs are smaller than those found for comparably
massive galaxies in the local Universe.

Importantly, our results are statistically inconsistent with those
derived for $z\approx$~2 radio galaxies and quasars, which suggested
that distant galaxies have SMBHs a factor of $\approx$~4--6 times more
massive than those found for comparably massive galaxies in the local
Universe (more than an order of magnitude higher than found here for
SMGs; e.g.,\ \citealt{mclure06}; \citealt{peng06}); see Fig.~1. One
factor in this discrepany may be related to the selection of the
source populations since high-$z$ radio galaxies and quasars by
definition have massive SMBHs and are likely to be comparatively
evolved objects (e.g.,\ \citealt{lauer07}); indeed, the X-ray luminous
subset of the broad-line SMGs would lie close to or above the local
SMBH--host galaxy relationship if it was assumed that they are hosted
by galaxy spheroids as massive as the typical SMG population (see
Fig.~1). However, as shown in \cite{alex08}, there is a key factor why
typical SMGs do not lie on the $M_{\rm BH}$--$M_{\rm SPH}$
relationship found for $z\approx$~2 quasars and radio galaxies: they
would overproduce the mass density of the most massive SMBHs in the
local Universe.

\section{Why SMGs cannot lie on the Elevated Black-Hole--Spheroid Mass Relationship Found for $z\approx$~2 Quasars and Radio Galaxies}

If we assume that SMGs lie a factor $\approx$~4--6 above the local
$M_{\rm BH}$--$M_{\rm SPH}$ relationship, as found for $z\approx$~2
quasars and radio galaxies then, given a spheroid mass of
$10^{11}$~$M_{\odot}$, they would have to host black holes of $M_{\rm
BH}\approx10^9$~$M_{\odot}$. The space density of $M_{\rm
BH}=10^9$~$M_{\odot}$ SMBHs in the local Universe is
$\Phi\approx10^{-5}$~Mpc$^{-3}$, which is the same as the observed
space density of SMGs at $z\approx$~2. This scenario is therefore only
plausible if SMGs are continously bright at submm wavelengths and
never ``switch off'' over the peak $z=$~1.7--2.8 redshift range of
submm activity (i.e.,\ a duty cycle of $\approx$~1.5 Gyrs). In order
to stay bright at submm wavelengths and maintain such high
star-formation rates for such a long duration, gas masses an order of
magnitude larger than those found for SMGs would be required, which
would only be feasible if the gas reservoir of SMGs is continously
replenished. Furthermore, the ultimate stellar masses of SMGs after a
1.5~Gyr lifecycle of star formation and SMBH growth would be
$>10^{12}$~$M_{\odot}$, which would lead to even larger SMBH masses,
further compounding the SMBH space-density problem. Indeed, given the
measured gas masses of SMGs, the submm-bright lifetime of SMGs is more
likely to be $\approx$~100--300~Myrs (i.e.,\ at any given snapshot
only $\approx$~10\% of galaxies that will be SMGs over the peak
$z=$~1.7--2.8 redshift range will be bright at submm wavelengths),
which leads to a duty-cycle corrrected space density for SMGs that is
an order of magnitude larger than that found for SMBHs with $M_{\rm
BH}=10^9$~$M_{\odot}$ in the local Universe
($\Phi\approx10^{-4}$~Mpc$^{-3}$).

The SMBH space-density arguments alone show that typical SMGs cannot
host black holes with $M_{\rm BH}\approx10^9$~$M_{\odot}$. However, we
can consider an alternative scenario, where the spheroid masses of
SMGs are overestimated by an order of magnitude (i.e.,\ $M_{\rm
SPH}\approx10^{10}$~$M_{\odot}$), which would place SMGs a factor of
$\approx$~4--6 above the local relationship. This hypothetical
scenario appears unlikely since the large star-formation rates of SMGs
and the likely duration of submm-bright activity would lead to an
increase in stellar mass of $\approx10^{11}$~$M_{\odot}$. Therefore,
unless this additional stellar growth only occurs outside of the
spheroid region, the mass of the spheroid in a typical SMG is likely
to be $\gg10^{10}$~$M_{\odot}$ (i.e.,\ given the high star-formation
rates of SMGs it would take just $\approx10^7$~yrs to "build" a
stellar mass of $10^{10}$~$M_{\odot}$). Since the observed spatial
extent of the CO emission in SMGs is confined to a compact region with
an inferred gas density consistent with that expected to produce a
galaxy spheroid (e.g.,\ \citealt{bouche07}), this suggests that the
star formation is predominantly confined to the galaxy spheroid
region, making this scenario unlikely.

We can also determine the largest SMBH mass that SMGs could host and
then explore whether the inferred spheroid mass is plausible, under
the assumption that the $M_{\rm BH}$--$M_{\rm SPH}$ ratio at
$z\approx$~2 is a factor $\approx$~4--6 above the local
relationship. On the basis of the duty-cycle corrected space density
of SMGs ($\Phi\approx10^{-4}$~Mpc$^{-3}$), the largest possible SMBHs
that typical SMGs could host without overproducing the local SMBH
space density are $M_{\rm BH}\approx3\times10^8$~$M_{\odot}$, which
would imply that the spheroid in SMGs could not be larger than $M_{\rm
SPH}\approx$~(3--5)~$\times10^{10}$~$M_{\odot}$. Since we have
considered the largest possible SMBH mass for SMGs, we are clearly
taking the extreme assumption that no significant SMBH growth will
occur in SMGs between $z\approx$~2 and the present day. Furthermore,
this spheroid mass constraint is consistent with the average gas mass
of SMGs and would therefore dictate that the spheroid can {\it only}
significantly grow during the SMG phase. These scenarios place severe
constraints on how much SMGs and their progeny can evolve and
therefore appear unlikely. Although here we have considered the
largest SMBH that an SMG can host, we note that significantly smaller
SMBHs would lead to spheroid masses of order
$\approx10^{10}$~$M_{\odot}$ (assuming that the $M_{\rm BH}$--$M_{\rm
SPH}$ ratio is a factor $\approx$~4--6 above the local relationship),
which appears unlikely unless the star formation in SMGs is
predominantly occuring outside of the spheroid region; see constraints
above. We note that \cite{hopkins06} took a similar approach to
estimate the general degree of evolution in the $M_{\rm BH}$--$M_{\rm
SPH}$ ratio, given space-density constraints in the local Universe,
also finding that typical $z\approx2$ systems cannot lie far from the
local relationship.

 \setcounter{figure}{1}
 \begin{figure}[!t]
 \plotone{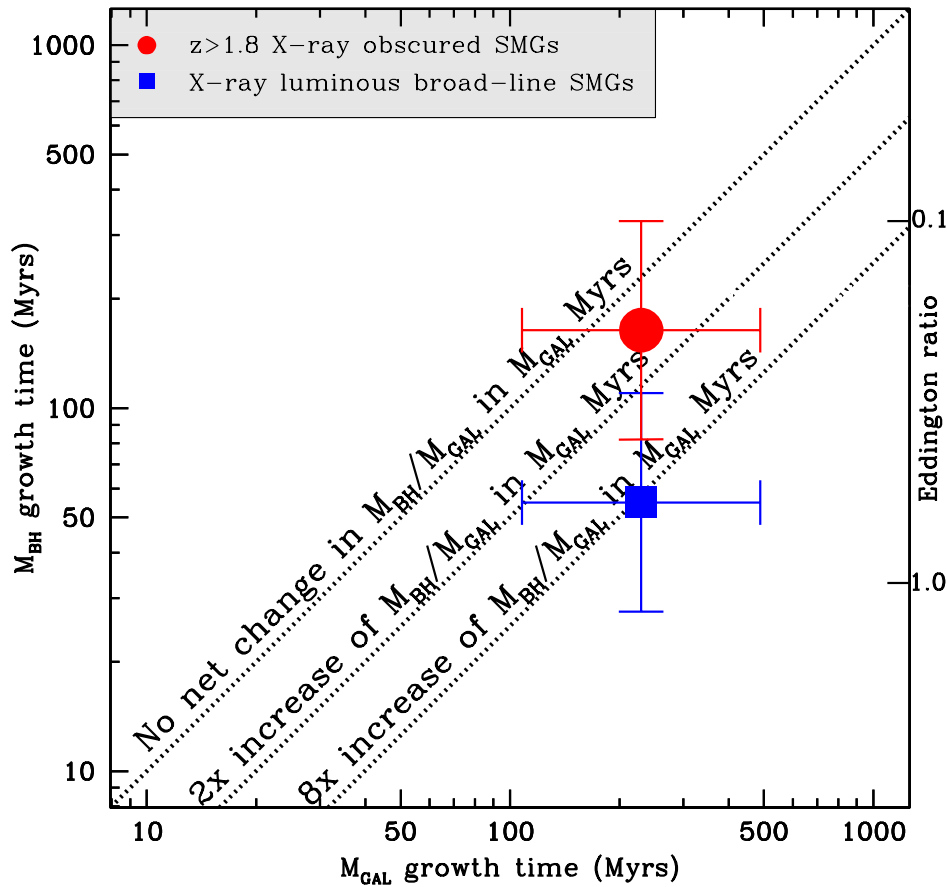}
 \caption{Growth times (i.e.,\ the time it would take to double the
   current mass) of the SMBH and host galaxy (spheroid) for SMGs
   (solid circle) and the X-ray luminous subset of the broad-line SMG
   population (solid square). Although very uncertain, the SMBH--host
   galaxy growth of SMGs is consistent with no net change in their
   $M_{\rm BH}$--$M_{\rm GAL}$ ratios for likely submm-bright
   lifetimes and a more AGN-dominated growth phase would be required
   to significantly move SMGs and their progeny up the $M_{\rm
   BH}$--$M_{\rm GAL}$ plane. The X-ray luminous subset of the
   broad-line SMG population may represent this AGN-dominated phase;
   see also Fig.~3 and \S5.2 of \cite{alex08} and \cite{copp08}.}
 \end{figure}

\section{How far can SMGs move up the Black-Hole--Spheroid Mass Plane?}

Due to the large space density of SMGs and their rapid growth rates,
our constraints show that SMGs and their progeny cannot lie the factor
$\approx$~4--6 above the local relationship found for $z\approx$~2
quasars and radio galaxies. However, our constraints do not rule out
the possibility that the SMBHs and spheroids in SMGs will grow beyond
our currently derived $M_{\rm BH}$--$M_{\rm SPH}$ ratio, although they
do provide tight limits. As shown above, on the basis of the
space-density constraints of SMGs at $z\approx$~2, the ultimate SMBH
mass of typical SMGs cannot exceed $M_{\rm
BH}\approx3\times10^8$~$M_{\odot}$, limiting the amount of SMBH growth
in SMGs and their progeny by up-to a factor of $\approx$~6.

We can explore how quickly SMGs can move up the $M_{\rm BH}$--$M_{\rm
SPH}$ plane by estimating the growth rates of the SMBH and host
galaxy; see Fig.~2. Although highly uncertain due to a number of
assumptions required to determine key parameters, the relative growth
of the SMBH and host galaxy appear to be similar, indicating that SMGs
will not significantly move up the $M_{\rm BH}$--$M_{\rm SPH}$ plane
over the likely duration of the submm-bright phase. For example, it
would take $\approx$~400~Myrs for the SMBH to grow by the maximum
factor of $\approx$~6 allowed. These analyses suggest that an
AGN-dominated growth phase, where the SMBH grows substantially faster
than the host galaxy, is required to significantly move SMGs and their
progeny up the SMBH--spheroid mass plane. An AGN-dominated growth
phase may be associated with optically bright quasars, where the
relative SMBH growth could exceed that of the galaxy spheroid. Indeed,
the relative SMBH--host galaxy growth times of the X-ray luminous
subset of the broad-line SMGs is sufficient to increase the
SMBH--spheroid mass ratio by a factor of $\approx$~8 over 200~Myrs;
see Fig.~2 and \S5.2 of \cite{alex08}. We note that the CO-derived gas
and host-galaxy masses of submm-detected optically luminous quasars
are consistent with those expected for an AGN-dominated phase that
follows the rapid SMBH--spheroid growth phase of luminous SMGs
\cite{copp08}, providing evidence for this evolutionary scenario.

We finally note that given the rapid growth rates of SMGs (i.e.,\
star-formation rates high enough to build a massive spheroid of
$M_{\rm SPH}\approx10^{11}$~$M_{\odot}$ in 100~Myrs), it is quite
amazing that the $M_{\rm BH}$--$M_{\rm SPH}$ ratio does not deviate
far from that seen in massive galaxies in the local Universe. This
suggests that some mechanism connects and regulates the growth of both
components.

\acknowledgements %%% Text of acknowledgements runs on after this command.

I thank the Royal Society for generous support and the following
collaborators for allowing me to present this research: A.~Blain,
F.~Bauer, N.~Brandt, S.~Chapman, K.~Coppin, R.~Ivison,
K.~Men{\'e}ndez-Delmestre, I.~Smail, and M.~Swinbank.

%%% THE BIBLIOGRAPHY
%%%
%%% CONSULT SECTION 3 OF "INSTRUCTIONS FOR AUTHORS" FOR HOW TO USE NATBIB.
%%% AUTHORS ARE ENCOURAGED TO USE EITHER THE "THEBIBLIOGRAPY" ENVIRONMENT
%%% BY UNCOMMENTING (DELETING THE "%" SYMBOL) THE COMMANDS BELOW, OR BY
%%% USING THE BIBTEX ENVIRONMENT. TO FIND OUT WHICH IS APPLICABLE TO YOUR
%%% CONTRIBUTION, CONSULT THE VOLUME EDITORS FOR YOUR PROCEEDINGS.
%%%

\end{document}